\documentclass{article}
\usepackage{amsmath, amssymb, amsfonts}
\title{On a regular modified C-metric}  
\author{Hristu Culetu, \\Ovidius University, Department of Physics and Electronics, \\ Mamaia Avenue 124, 900527 Constanta, Romania, \\e-mail : hculetu@yahoo.com}

\begin{document}
\numberwithin{equation}{section}
\pagenumbering{arabic}
\maketitle
\newcommand{\fv}{\boldsymbol{f}}
\newcommand{\tv}{\boldsymbol{t}}
\newcommand{\gv}{\boldsymbol{g}}
\newcommand{\OV}{\boldsymbol{O}}
\newcommand{\wv}{\boldsymbol{w}}
\newcommand{\WV}{\boldsymbol{W}}
\newcommand{\NV}{\boldsymbol{N}}
\newcommand{\hv}{\boldsymbol{h}}
\newcommand{\yv}{\boldsymbol{y}}
\newcommand{\RE}{\textrm{Re}}
\newcommand{\IM}{\textrm{Im}}
\newcommand{\rot}{\textrm{rot}}
\newcommand{\dv}{\boldsymbol{d}}
\newcommand{\grad}{\textrm{grad}}
\newcommand{\Tr}{\textrm{Tr}}
\newcommand{\ua}{\uparrow}
\newcommand{\da}{\downarrow}
\newcommand{\ct}{\textrm{const}}
\newcommand{\xv}{\boldsymbol{x}}
\newcommand{\mv}{\boldsymbol{m}}
\newcommand{\rv}{\boldsymbol{r}}
\newcommand{\kv}{\boldsymbol{k}}
\newcommand{\VE}{\boldsymbol{V}}
\newcommand{\sv}{\boldsymbol{s}}
\newcommand{\RV}{\boldsymbol{R}}
\newcommand{\pv}{\boldsymbol{p}}
\newcommand{\PV}{\boldsymbol{P}}
\newcommand{\EV}{\boldsymbol{E}}
\newcommand{\DV}{\boldsymbol{D}}
\newcommand{\BV}{\boldsymbol{B}}
\newcommand{\HV}{\boldsymbol{H}}
\newcommand{\MV}{\boldsymbol{M}}
\newcommand{\be}{\begin{equation}}
\newcommand{\ee}{\end{equation}}
\newcommand{\ba}{\begin{eqnarray}}
\newcommand{\ea}{\end{eqnarray}}
\newcommand{\bq}{\begin{eqnarray*}}
\newcommand{\eq}{\end{eqnarray*}}
\newcommand{\pa}{\partial}
\newcommand{\f}{\frac}
\newcommand{\FV}{\boldsymbol{F}}
\newcommand{\ve}{\boldsymbol{v}}
\newcommand{\AV}{\boldsymbol{A}}
\newcommand{\jv}{\boldsymbol{j}}
\newcommand{\LV}{\boldsymbol{L}}
\newcommand{\SV}{\boldsymbol{S}}
\newcommand{\av}{\boldsymbol{a}}
\newcommand{\qv}{\boldsymbol{q}}
\newcommand{\QV}{\boldsymbol{Q}}
\newcommand{\ev}{\boldsymbol{e}}
\newcommand{\uv}{\boldsymbol{u}}
\newcommand{\KV}{\boldsymbol{K}}
\newcommand{\ro}{\boldsymbol{\rho}}
\newcommand{\si}{\boldsymbol{\sigma}}
\newcommand{\thv}{\boldsymbol{\theta}}
\newcommand{\bv}{\boldsymbol{b}}
\newcommand{\JV}{\boldsymbol{J}}
\newcommand{\nv}{\boldsymbol{n}}
\newcommand{\lv}{\boldsymbol{l}}
\newcommand{\om}{\boldsymbol{\omega}}
\newcommand{\Om}{\boldsymbol{\Omega}}
\newcommand{\Piv}{\boldsymbol{\Pi}}
\newcommand{\UV}{\boldsymbol{U}}
\newcommand{\iv}{\boldsymbol{i}}
\newcommand{\nuv}{\boldsymbol{\nu}}
\newcommand{\muv}{\boldsymbol{\mu}}
\newcommand{\lm}{\boldsymbol{\lambda}}
\newcommand{\Lm}{\boldsymbol{\Lambda}}
\newcommand{\opsi}{\overline{\psi}}
\renewcommand{\tan}{\textrm{tg}}
\renewcommand{\cot}{\textrm{ctg}}
\renewcommand{\sinh}{\textrm{sh}}
\renewcommand{\cosh}{\textrm{ch}}
\renewcommand{\tanh}{\textrm{th}}
\renewcommand{\coth}{\textrm{cth}}

\begin{abstract}
A particular form of the C-metric is investigated, giving it a non-standard interpretation and removing any singularity at $r = 0$. In the weak field limit of the accelerating black hole, the proper acceleration $A$ of a static observer is constant and the geometry becomes conformally-flat (anti de Sitter). The stress tensor is of $\Lambda$-type ($\Lambda = -3a^{2}/8\pi G$) and its energy density is negative. We propose that $\Lambda$ is responsible of inertial forces that appear in uniformly accelerated systems (far from the accelerating source $m$ and for $r << 1/a$ the dominant term in the expression of $a^{r}$ is $-a~cos\theta$). The components of the stress tensor and all invariants of the conformally-flat Schwarzschild spacetime are regulated by means of the exponential factor $exp(-k/r),~ k > 0$.
 \end{abstract}
 
 \section{Introduction}
 The well-known C-metric belongs to a class of spacetimes with boost-rotation symmetries \cite{BS, PG, JP, AG, RM}. Kinnersley and Walker \cite{KW} have shown since 1970 that the C-metric describes a pair of uniformly accelerated black holes (BHs) in the Minkowski spacetime. Their acceleration is rooted from the conical singularities produced by a strut between the two BHs or two semi-infinite strings connecting them to infinity. The pair creation of BHs may be possible in a background with a cosmological constant $\Lambda$ as this supplies the negative potential energy \cite{PD, MR}.
 
 To find the physical interpretation of the $\Lambda \neq 0$ case, Podolsky and Griffiths \cite{PG} introduced a new coordinate system adapted to the motion of two uniformly accelerating test particles in de Sitter (dS) space. However, the curvature singularity at $r = 0$ is also present, as in the previous papers on the subject. A physical meaning of the C-metric with a negative $\Lambda$ was given by Podolsky \cite{JP}. He showed that this exact solution of Einstein's field equations describes uniformly accelerated BHs in anti de Sitter (AdS) universe, using a convenient coordinate system. Recently Arnfinnsson and Gron \cite{AG} found a new source (a singular accelerating mass shell) of the C-metric, using the Israel junction conditions. They took advantage of the C-metric in spherical coordinates, previously used in \cite{GKP}. The shell consists of a perfect fluid that creates a jump of the extrinsic curvature when the shell is crossed. 
 
 Our purpose in this work is to develop the previous studies on the C-metric, giving it a different interpretation and ruling out any singularity at $r = 0$. We get rid of the conical singularity by neglecting $2amG/c^{4}$ w.r.t. unity ($a$ is the constant acceleration and $m$ is the BH mass). Our basic geometry is conformal to the Schwarzschild (KS) metric and possesses axial symmetry. We found that the const $a$ is the proper acceleration of a static observer and the spacetime is AdS far from the mass $m$, with a cosmological constant $\Lambda = -3a^{2}$. To get rid of the curvature singularity at $r = 0$ we make use of the modified KS geometry \cite{HC1, HC2} so that the spacetime becomes AdS both when $r\rightarrow 0$ and when $r \rightarrow \infty$.
 
 Throughout the paper we use geometrical units $G = c = 1$, unless otherwise specified.
 
 \section{Conformal Schwarzschild metric}
 To begin with, we write down the C-metric in spherical coordinates, following \cite{AG, GKP} 
    \begin{equation}
  ds^{2} = \frac{1}{(1 + ar cos\theta)^{2}} \left(- Q dt^{2} + \frac{1}{Q} dr^{2} + \frac{r^{2}}{P}d\theta^{2} + Pr^{2}sin^{2}\theta d\phi^{2}\right),
 \label{2.1}
 \end{equation}
 where $Q = (1 - a^{2}r^{2})(1 - 2m/r), P = 1 + 2amcos\theta$, $a$ is a constant amd $m$ is the BH (or a point particle) mass. According to the authors of \cite{AG} and \cite{GKP}, Eq. (2.1) can be viewed as a nonlinear combination of the KS and Rindler geometries, representing the metric around an accelerating point particle or a BH. The function $P$ is related to a conical singularity along the symmetry axis where the sources of the acceleration $a$ (a cosmic string or a strut) are supposed to be located.
 
 Let us observe that the term $2amcos\theta$ in the expression of $P$ is very low compared to unity when reasonable values for $a$ and $m$ are used. We have, indeed, $am = amG/c^{4} = ma/(c^{4}/G)$, where $c^{4}/G \approx 10^{44} N$ is a maximal force. Therefore, we shell work in the approximation $P = 1$. Something similar is valid for the function $Q$. The term $a^{2}r^{2}/c^{4}$ contains the factor $1/c^{4} \approx 10^{-34} s^{4}/m^{4}$ and, therefore, we take $Q = 1 - 2m/r$ from now on, but we will preserve the first order terms $ar/c^{2}$ and $2Gm/c^{2}r$. In other words, our areas of study are $am << 1$ and $a^{2}r^{2} << 1$. The condition $a^{2}r^{2} << 1$ is compatible with the removal of the singularity at $r = 0$, that will be done in Section 4.
 
 The above approximations yield the conformal KS metric
     \begin{equation}
  ds^{2} = \frac{1}{(1 + ar cos\theta)^{2}} \left[- (1 - \frac{2m}{r}) dt^{2} + \frac{dr^{2}}{1 - \frac{2m}{r}}  + r^{2} d \Omega^{2}\right] ,
 \label{2.2}
 \end{equation}
 where $d \Omega^{2}$ stands for the metric on the unit 2-sphere. The spacetime (2.2) has a Bh horizon at $r = 2m$. We might consider (2.2) as a metric by itself and forget that it is rooted from the C-metric (the constant $C$ related to the range of the $\phi$-coordinate plays no role here because $\phi \in (0, 2\pi)$). The Ricci scalar for (2.2) is given by
      \begin{equation}
      R^{a}_{~a} = -12a^{2} - \frac{12am}{r^{2}}cos\theta ~(1 - ar cos\theta)
 \label{2.3}
 \end{equation} 
 It is worth noticing that there is a curvature singularity when $r \rightarrow 0$ and asymptotically $ R^{a}_{~a} = -12a^{2}$. The singularity will be removed in Section 4, after a slight modification of the geometry (2.2).
 
 Let us take now a static observer in the geometry (2.2) with the velocity vector field
 \begin{equation}
 u^{b} = \left(\frac{1 + ar cos\theta)}{\sqrt{1 - \frac{2m}{r}}}, 0, 0, 0\right), 
 \label{2.4}
 \end{equation} 
where $b$ labels $(t, r, \theta, \phi)$. The acceleration 4-vector $a^{b} = u^{a}\nabla_{a}u^{b}$ has the nonzero components
 \begin{equation}
 a^{r} = (\frac{m}{r^{2}} - acos\theta + \frac{3am}{r}cos\theta)(1 + ar cos\theta),~~~ra^{\theta} = asin\theta ~(1 + ar cos\theta)
 \label{2.5}
 \end{equation} 
 with the proper acceleration
  \begin{equation}
  A \equiv \sqrt{a^{b}a_{~b}} = \sqrt{\frac{\left[acos\theta (1 - \frac{2m}{r}) - \frac{m}{r^{2}}(1 + ar cos\theta)\right]^{2} + (1 - \frac{2m}{r}) a^{2}sin^{2}\theta }{1 - \frac{2m}{r}}}
 \label{2.6}
 \end{equation} 
 In the weak field limit ($(m = 0), A = a$ and the interpretation of the constant $a$ as the acceleration of a static observer is justified. In addition, the horizon surface gravity is given by
   \begin{equation}
  \kappa = \sqrt{a^{b}{a_{b}}}~ \sqrt{-g_{tt}}|_{r = 2m} = \frac{m}{r^{2}}|_{r = 2m} = \frac{1}{4m},
 \label{2.7}
 \end{equation} 
 exactly as for a KS black hole. It does not depend on $\theta$, as it should be.
 
 In the expression of $a^{r}$ from (2.5) we observe that the dominant term is $-acos\theta$, when $r >> 2m, ar << 1$ and $ma << 1$. If we consider a test particle of mass $m'$, located on the $z$-axis ($\theta = 0$), we have in the above conditions $m' ra^{\theta} = 0$ and $m' a^{r} \approx -m'a$ (the  term $mm'/r^{2}$ represents the Newtonian gravitational force from the classical mechanics and $-m'a$ is the inertial force; because of the Newton's constant $G$, the Newtonian force is obviously much weaker than the inertial force). The standard view asserts that the inertial forces originate from a fictitious field but here the field comes from the nonflat character of the geometry (even when $m = 0$): the nonzero Ricci scalar  $R^{a}_{~a} = -12a^{2}$ or the Kretschmann invariant $K = 24a^{4}$. Therefore, in our opinion, the inertial forces which appear in accelerated systems originates from the negative cosmological constant generated by acceleration (when $m = 0$, the spacetime (2.2) is AdS, with $\Lambda = - 3a^{2}$). A similar dependence of $\Lambda$ on the acceleration $a$ has been reached by Cvetic and Griffies \cite{CG} in their studies upon domain walls with an AdS geometry on one side (see also \cite{CGS}). Moreover, they showed that asymptotically in the AdS spacetime, geodesic particles moving in the $z = rcos\theta$ direction are Rindler particles whose constant proper acceleration is proportional to $\sqrt{- \Lambda}$. In conclusion, the spacetime in an accelerated coordinate system turns out not to be flat but conformally flat (AdS) and the metric cannot be obtained by a coordinate transformation from the Minkowski space. The curvature is created by the agent who accelerates the test particle. That remind us the quantum Unruh effect - the heat bath felt by an accelerating observer. We must keep in mind that the thermal bath of particles - which is an intermediary - has energy and, as a consequence, creates curvature (by the agent who put the reference system in accelerated motion). This curvature is, however, neglected (the thermal bath has a negligible temperature) and the metric remains flat. 
 
 \section{Anisotropic stress-energy tensor}
  We look now for the sources of the spacetime (2.2), namely the stress tensor to lie on the r.h.s. of Einsteins' equations $G_{ab} = 8\pi T_{ab}$ in order that (2.2) to be an exact solution. 
 By means of the software package Maple - GRTensorII, one finds that
   \begin{equation} 
   \begin{split} 
  T^{t}_{~t} = -\rho = \frac{3a^{2}}{8\pi}\left(1 + \frac{2m}{ar^{2}}cos\theta \right) ,~~~ T^{r}_{~r} = p_{r} = - \rho,\\ T^{\theta}_{~\theta} = T^{\phi}_{~\phi} = p_{\theta} = p_{\phi} =  \frac{3a^{2}}{8\pi} \left(1 - \frac{2m}{r}cos^{2}\theta \right)  
\label{3.1}
\end{split}
\end{equation} 
Because $T^{t}_{~t} =  T^{r}_{~r}$, and $T^{\theta}_{~\theta} = T^{\phi}_{~\phi}$, the spacetime has an infinite set of comoving reference frames \cite{ID1, ID2, DK}. One sees also that $a \rightarrow 0$ leads to a vanishing $ T^{a}_{~b}$, even though $m \neq 0$. This is because the metric (2.2) becomes the KS metric which is a empty space solution of the Einstein equations. Moreover, the weak field limit ($m = 0$) gives us an AdS spacetime, with $\Lambda = -3a^{2}$, when the anisotropic fluid becomes isotropic, with
 \begin{equation}  
   -\rho = \frac{3a^{2}}{8\pi} = p_{r} =  p_{\theta}.
\label{3.2}
\end{equation} 
 In this case $ T^{a}_{~b}$ is Lorentz-invariant ($ T^{a}_{~b} \propto \Lambda \delta^{a}_{b}$). The energy density of the gravitational fluid is negative and so the weak energy condition is not observed. However, when $m \neq 0$, for small $r$ and with $cos\theta < 0$ or $a < 0$, one might obtain $\rho > 0$.
 
 With all the fundamental constants introduced in (3.1), the energy density looks like
    \begin{equation}  
   \rho = -\frac{3a^{2}}{8\pi G}\left(1 + \frac{2Gm}{ar^{2}}cos\theta \right) = -\frac{3a^{2}}{8\pi G}\left(1 + \frac{2g}{a}cos\theta \right)
\label{3.3}
\end{equation} 
 where $g = Gm/r^{2}$ is the Newtonian acceleration due to the mass $m$, at the distance $r$ from the point particle or the BH. If we take, for example, $m = 1 Kg, ~r = 1m$ and $a = 1m/s^{2}$, one obtains $g << a$ and $\rho \approx -3a^{2}/8\pi G = -10^{10} ergs/cm^{3}$. Consequently, the effect of the acceleration $a$ of the mass $m$ is much greater than that of the gravitational field $g$ due to the mass $m$. We also note that $\rho \propto a^{2}$, as in the Newtonian gravitation. 
 
 For the tangential pressures we may perform a similar analysis
   \begin{equation}  
   p_{\theta} = p_{\phi} =  \frac{3a^{2}}{8\pi G} \left(1 - \frac{2Gm}{rc^{2}}cos^{2}\theta \right) = \frac{3a^{2}}{8\pi G} \left(1 - \frac{r_{g}}{r}cos^{2}\theta \right)  
\label{3.4}
\end{equation}  
 where $r_{g}$ is the gravitational radius of the mass $m$. Outside the source (if the particle is not a BH) we have $r > r_{g}$ and, therefore, $p_{\theta} > 0$ for any $\theta \in (0,\pi)$. All the components of the stress tensor are, of course, divergent at $r = 0$ because of the curvature singularity at the origin of coordinates.
 
\section{Regular stress-energy tensor}
We try now to modify the metric (2.2) in order to render all the physical parameters finite, both at the origin and at infinity. For that purpose we apply to a previous choice \cite {HC1, HC2} and introduce the factor $exp(-k/r), k > 0$ as a cutoff in the KS metric. Therefore, the geometry (2.2) becomes
     \begin{equation}
   ds^{2} = \frac{1}{(1 + ar~ cos\theta)^{2}} \left[-\left(1 - \frac{2m}{r} e^{-\frac{k}{r}}\right) dt^{2} + \frac{1}{1 - \frac{2m}{r} e^{-\frac{k}{r}}} dr^{2} + r^{2} d \Omega^{2}\right].         
 \label{4.1}
 \end{equation}
A discussion of the geometry inside the square parantheses from (4.1) in terms of the values of $k$ has been given in \cite{HC2} and will not be repeated here. The exponential factor makes, indeed, finite all invariants corresponding to the metric (4.1) and all the components of the stress tensor. The curvature scalar has the form
     \begin{equation}
	 R^{a}_{a} = -12a^{2} + \frac{2}{r^{5}} (...) e^{-\frac{k}{r}}	
 \label{4.2}
 \end{equation}
where (...) stands for a fourth degree polinomial of $r$. It is clear from (4.2) that the Ricci scalar tends to the constant value $-12a^{2}$ when $r \rightarrow 0$ or when $r \rightarrow \infty$. 

As far as  $T^{a}_{~b}$ is concerned, we obtain the following expressions for its components
 \begin{equation} 
 \begin{split}
T^{t}_{~t} = -\rho = \frac{3a^{2}}{8\pi} - \left[\frac{q^{2}}{8\pi r^{4}} (1 + ar~cos\theta) - \frac{3am}{4\pi r^{2}} cos\theta \right]e^{-\frac{q^{2}}{2mr}}, \\ T^{r}_{~r} = p_{r} = - \rho,~~~ T^{\theta}_{~\theta} = T^{\phi}_{~\phi} = \frac{3a^{2}}{8\pi} -\\
	\left[\frac{3ma^{2}}{4\pi r} cos^{2}\theta - \frac{q^{2}}{8\pi r^{4}} (1 + ar~cos\theta) (1 + 3ar~cos^{2}\theta) 
	+ \frac{q^{4}}{32\pi mr^{5}} (1 + ar~cos\theta)^{2}\right] e^{-\frac{q^{2}}{2mr}}
\end{split}
\label{4.3}
\end{equation} 
where the positive constant $k$ has been replaced with $q^{2}/2m$ for a particle of charge $q$ (see \cite{HC2}). We see from (4.3) that the mass $m$ and the charge $q$ play no role when $r \rightarrow 0$ or  $r \rightarrow \infty$, due to the exponential cutoff. What remains is the first term proportional to the cosmological constant $\Lambda = - 3a^{2}$, i.e. the values of $\rho, p_{r}$ and $p_{\theta}$ given by (3.3) when $m = 0$. 

A similar effect takes place for the components of the acceleration of a static observer. We have
   \begin{equation} 
   \begin{split}
 a^{r} = -acos\theta ~(1 + ar cos\theta) + \\  \left[\frac{m}{r^{2}}(1 + ar cos\theta)(1 + 3ar cos\theta) -  \frac{q^{2}}{2r^{3}} (1 + ar cos\theta)^{2}\right] e^{-\frac{q^{2}}{2mr}}  
 \end{split} 
\label{4.4}
\end{equation} 
and
   \begin{equation} 
    ra^{\theta} = asin\theta ~(1 + ar cos\theta),~~~a^{t} = a^{\phi} = 0.	
\label{4.5}
\end{equation}
We observe that, at $r = 0, ~a^{r} = -acos\theta$ and $ ra^{\theta} = asin\theta$, whence $\sqrt{a^{b}_{~b}}|_{r = 0} = A_{0} = a$. However, for an asymptotic, static observer, both nonzero components are divergent. 

Let us remark that, with $P \neq 1, Q \neq 1$ in (2.1), the Ricci scalar and all nonzero components of $T^{a}_{~b}$ are not regulated at $r = 0$ even when the exponential factor is used because there are terms in their expressions which do not contain $exp(-k/r)$.

\section{Conclusions}
Our main assumption in this paper refers to the non-flat character of the metric in an accelerating reference system. Far from a uniformly accelerated BH (or a point particle) of mass $m$ the metric seems to be not flat but conformally-flat, with axial symmetry. The associated energy-momentum tensor is of $\Lambda$-type ($\Lambda = -3a^{2}/8\pi G$), Lorentz invariant, with negative energy density and the Ricci scalar $R^{a}_{~a} = -12a^{2}$. 

We consider that $\Lambda$ is responsible for the inertial forces that give rise in accelerated reference systems. Therefore, in our view the metric in an accelerating system cannot be obtained from the Minkowski metric through a coordinate transformation.

\end{document}